# Dispersion-managed soliton fiber laser with random dispersion, multiphoton absorption and gain dispersion

Gurkirpal Singh Parmar[1], Rajib Pradhan[2], B.A. Malomed[3,4] and Soumendu Jana[1*]

[1]School of Physics and Materials Science, Thapar Institute of Engineering and Technology, Patiala, India
[2]Department of Physics, Midnapore College, Midnapore, India
[3]Department of Physical Electronics, School of Electrical Engineering, Faculty of Engineering, Tel Aviv University, P.O.B. 39040, Tel Aviv, Israel
[4]Center for Light-Matter Interaction, Tel Aviv University, P.O.B. 39040, Tel Aviv, Israel

E-mail: soumendujana@yahoo.com

## Abstract

We address the generation and interaction of dispersion-managed dissipative solitons (DMDS) in a model of fiber lasers with the cubic-quintic nonlinearity, multiphoton absorption and gain dispersion. Both anomalous and normal segments of the dispersion map include random dispersion fluctuations. Effects of the gain dispersion, higher-order nonlinearity and randomness on the generation of DMDS are demonstrated. The solitons exhibit breather-like evolution, and are found to be robust, up to a certain critical level of the random-dispersion component, which is sufficiently high. The roles of multiphoton absorption, gain dispersion and nonlinearity on the DMDS are also identified in the absence of randomness. Pair wise interactions of solitons lead, most typically, to their merger, with breaking of the left-right symmetry. The outcome of the collisions is more sensitive to the initial temporal separation between the solitons than to their phase difference.

Keywords: dispersion- managed soliton, fiber laser, multiphoton absorption, dissipative soliton, random dispersion

## 1. Introduction

Nowadays, fiber lasers hold the leading position in industrial and medical applications due to their compact size yet high output powers, lower cost yet high-quality of produced beams, structural simplicity, and robustness [1-4]. Fiber lasers are naturally compatible with fiber-optic telecoms links and data-processing schemes. Their electrical-to-optical efficiency is remarkable high – typically, ≈ 50% in commercial applications, exceeding 70% in laboratory experiments, and the optical-to-optical conversion efficiency as high as 70% may be achieved [5, 6].

The laser is basically built of a piece of a fiber doped by rare-earth elements, such as erbium or ytterbium, which is confined by mirrors (generally, these are fiber Bragg gratings) to form the cavity structure. The fiber cavity can be either end-pumped by one or more external laser beams (usually produced by a diode or by other sources), or side-pumped by many beams. A large variety of core structures dopants, and overall configurations enable diverse applications, helping the fiber lasers to compete with lasers of other types.

The performance of a conventional fiber laser can be significantly improved by enabling it to emit soliton pulses, which may be achieved with the help of several self-started, passive mode-locking techniques [7,8], which are based on effective nonlinear amplifiers, that, in turn, often make use of saturable absorbers. Owing to its easier implementation and comprehensiveness, nonlinear polarization rotation [9] is frequently preferred for the implementation of the saturable absorption. Alternatively, nonlinear loop mirror (NOLM) is used too [10]. Other schemes for nonlinear amplification were theoretically elaborated too, including those using the second-harmonic generation [11] and dual-core fibers [12]. Nonlinear-polarization-rotating cavities always produce scalar solitons, due to the polarization-dependent character of their operation [13]. To produce a vector (two-component) soliton, the polarization dependence should be removed as much as possible from the operation regime. In this context, semiconductor saturable absorber mirror (SESAM) [14, 15] and currently popular carbon nanotubes are extensively used for the mode locking [16]. In addition to the carbon nanotubes, a variety of two-dimensional optical materials, including, graphene and molybdenum disulfide ($MoS_2$), owing to their fast recovery, higher absorption power and higher saturation fluency, are gradually replacing SESAM [17-19].

The nonlinearity (here saturable) not only creates pulses from noise but also sharpens them. But excess nonlinearity, that cannot be compensated by the fiber's dispersion, gives rise to detrimental nonlinear phase modulation and, eventually, distortion or fragmentation of the pulse. The dispersion management (DM) technique can remarkably improve the operation of fiber lasers by controlling the excess nonlinearity effect [20-22]. Similar to the DM in transmission lines [23, 24], DM fiber lasers too use the group-velocity-dispersion (GVD) map that alternates periodically between the anomalous and normal values, keeping the net average GVD positive, negative, or close to zero [25]. Adding the DM to fiber-laser cavities, on the one hand, drastically reduces the detrimental effect of the noise induced by the amplified spontaneous emission, and, on the other hand, greatly extends the power range of the generation of single pulses [26]. Thus, in an actively mode-locked dispersion-managed fiber ring laser, the strong DM reduces the detrimental time jitter. The location of the filter, amplifier and modulator in the cavity significantly affects the magnitude of the time and energy jitter [23]. Further, bound states of DM solitons can be obtained at nearly-zero net cavity GVD in a passively mode-locked erbium-doped fiber ring laser [27]. Passively mode-locked polarization-maintaining, figure-of-eight erbium-doped fiber lasers with a DM cavity can generate extremely robust pulses [28]. Yb-doped fiber lasers generate anti-symmetric DM solitons (tightly bound soliton pairs with the $\pi$ phase difference) by virtue of a strong- DM map [29]. It has also been demonstrated that the passive mode-locking in an Er-doped fiber laser with an embedded polyimide film containing dispersed single-wall carbon nanotubes generates a high-power ultrashort-pulse output with a small irradiation power absorbed by the film [30]. A review on the steady-state pulse dynamics through cavity dispersion engineering has been presented in [31] op.cit., which discuss the characteristic features as well as the stabilization mechanism of average solitons, DM solitons, dissipative solitons, giant-chirped pulses and similaritons.

Light transmission in any real fiber, hence the operation of DM fiber laser cavities too, give rise to linear and nonlinear losses [32, 33]. Therefore, an appropriate gain must be introduced in the system to support persistent solitons. The resulting pulses are often called dissipative solitons, being maintained by the gain-loss balance in addition to the GVD –self-phase modulation balance [34, 35]. The operation of a fiber laser on the wavelength of 2 µm in the regime of generation of DM dissipative solitons (DMDS) has been demonstrated [36]. Both stretched and positively chirped DMDSs aware demonstrated in mode-locked lasers [37]. Using a strong -DM cavity, a harmonically mode-locked Er-fiber soliton lasers can be built. DMDSs may be stabilized even in presence of the detrimental effect of the stimulated Raman scattering [38]. The gain-guided soliton-operation regime of DM fiber lasers has been demonstrated both experimentally and in numerical simulations [39]. Besides the anomalous GVD, the dispersion with the normal sign may also support dissipative solitons [40]. The all-normal GVD has been used for the trapping of dissipative solitons in fiber laser [41], as well as for the generation of dissipative vector solitons in a dispersion-managed fiber laser [42]. Insertion of chirped fiber Bragg grating generates net normal cavity GVD in the all-anomalous-dispersion fiber. In spite of the large frequency chirp of the dissipative soliton formed in the cavity, the polarization rotation, as well as polarization-locked dissipative vector solitons may be produced in a robust form [43].

Although soliton fiber lasers have been widely investigated, some of relevant issues were not addressed yet. In particular, the role of random dispersion and multiphoton absorption/emission has not been addressed explicitly. Most of the theoretical modelling considered ideal fibers with fixed core diameter and doping concentration. In reality, real fibers often include various imperfections due to fluctuations of the cross-section area, dopant concentration, inhomogeneities of the refractive index, bending, ellipticity and external stress, etc. [44, 45]. These imperfections may be random, manifesting themselves through random effective GVD. For real fibers, the random fluctuation in the core diameter may lie in the range of $\pm(3 \text{ to } 4)\%$ [46]. The influence of random GVD may be essential for sophisticated fiber-laser cavities [47, 48]. Another issue arises due to the ultrashort duration and high peak power of the output of the fiber laser. The high power may excite fifth-order nonlinear effects [49-52]. Although the quintic nonlinear term is weak in comparison with the basic cubic nonlinearity, the combined action of the self-focusing cubic and defocusing nonlinearities, which correspond, respectively, to real parts of the cubic and quintic susceptibilities, $\chi^3$ and $\chi^5$, gives rise to noteworthy pulse features and dynamics. The corresponding imaginary parts of $\chi^3$ and $\chi^5$ account for the two-photon absorption (TPA) and three-photon absorption (3PA) effects, respectively, in the fiber cavity [53, 54].

Further, the effect of the finite gain bandwidth becomes important for femtosecond and few-picosecond fiber-laser pulses [55].

In this work we investigate the role of the random-dispersion component of the DM map, gain dispersion, and multiphoton absorption in the generation and dynamics of DMDSs. The field dynamics in the lossy DM fiber-laser cavity with the cubic-quintic nonlinearity, gain dispersion and multiphoton absorption is modelled by the complex cubic-quintic Ginzburg-Landau equation (CQGLE) for scaled amplitude $E$ of the electromagnetic field:

$$i\frac{\partial E}{\partial z} + \frac{D(z)}{2}\frac{\partial^2 E}{\partial t^2} + |E|^2 E - \gamma|E|^4 E$$
$$= \frac{i}{2}(g_0 - \alpha)E + \frac{id}{2}\frac{\partial^2 E}{\partial t^2} - iK|E|^2 E - i\nu|E|^4 E, \quad (1)$$

where $z$ is the normalized propagation length and $t$ the retarded time, while $D(z)$ represents the DM map, which includes a small randomly varying small component. The third and fourth terms in Eq. (1) represent the cubic and quintic nonlinearities, respectively. If the intra-band relaxation time is much smaller than the temporal width of pulses governed by Eq. (1), the gain spectrum, $g(\omega)$, may be expanded in the Taylor series around carrier frequency $\omega_0$. Accordingly, the first two terms on the right-hand side of Eq. (1) represent the peak gain, $g_0$, and the gain dispersion, $d$, while $\alpha$ is the linear loss, which determines the net gain, $\Delta g = (g_0 - \alpha)$. The third and fourth terms on the right-hand side are the contributions of TPA and 3PA, respectively, with the respective scaled coefficients $K$ and $\nu$, which are related to the imaginary parts of $\chi^3$ and $\chi^5$ coefficients: $\alpha_2 = 3\omega \operatorname{Im}(\chi^{(3)})/2n_0^2 c^2 \varepsilon_0$, and $\alpha_3 = 5\omega \operatorname{Im}(\chi^{(5)})/2n_0^3 c^3 \varepsilon_0^2$.

It is relevant to mention that, formally speaking, the model based on Eq. (1) with $\Delta g > 0$ is unstable, as the zero background is made unstable by the effective linear gain. Nevertheless, the presence of the nonzero average dispersion makes this instability convective, as the emerging perturbations are carried away by their GVD-induced group velocity from the DMDS, and are gradually attenuated by the dispersive losses. Due to this mechanism, soliton-like pulses become effectively stable, cf. a similar situation reported in [12].

To handle Eq. (1), we use the variational approximation (VA). Generally, VA is valid when the operator which corresponds to the evolution equation (1) is self-conjugated. Obviously, the right-hand side of Eq.(1) cannot be written in the self-conjugated form. Therefore, we adopt VA in conjugation with the Rayleigh's dissipative function. The VA explicitly predicts evolution of individual parameters of the optical pulses, which we approximated by the sech ansatz:

$$E(z,t) = A(z)\operatorname{sech}\left(\frac{t}{W(z)}\right)\exp(i\phi(z)), \quad (2)$$

where $A(z)$, $W(z)$ and $\phi(z)$ are the complex amplitude and real temporal pulse width and phase, respectively. Following the standard VA procedure [54,56,57], the relation between the soliton's amplitude and width, together the respective evolution equations are derived, if the relatively weak gain and loss terms are not taken into regard, for the time being:

$$\frac{|A(z)|^2 W^2(z)}{D(z)} - \frac{16\gamma |A(z)|^4 W^2(z)}{15 D(z)} = 1. \quad (3)$$

$$\frac{dW(z)}{dz} = \left[-\frac{d}{3W}(1-\sqrt{M}) - \frac{2KW}{3s}(1-\sqrt{M})^2\right.$$
$$\left. - \frac{4\nu W}{15s^2}(1-\sqrt{M})^3 - \frac{2s}{W\sqrt{M}}\frac{dD}{dz}\right.$$
$$\left. + W(z)(g_0-\alpha)(1-\sqrt{M})\right]$$
$$/\left[-\frac{4sD}{W^2\sqrt{M}} + (1-\sqrt{M})\right], \quad (4)$$

$$\frac{dA(z)}{dz} = \left[\frac{1}{2AW^2 - 4sW^2 A^3}\right]\frac{dD}{dz}$$
$$-A(1-sA^2)\left[-\frac{d}{3W}(1-\sqrt{M}) - \frac{2KW}{3s}(1-\sqrt{M})^2\right.$$
$$\left. - \frac{4\nu W}{15s^2}(1-\sqrt{M})^3 - \frac{2s}{W\sqrt{M}}\frac{dD}{dz}\right.$$
$$\left. + W(z)(g_0-\alpha)(1-\sqrt{M})\right]$$
$$/\left(W(1-2sA^2)\left[-\frac{4sD}{W^2\sqrt{M}} + (1-\sqrt{M})\right]\right) \quad (5)$$

where $M \equiv 1 - 4sD/W^2$ and $s \equiv 16\gamma/15$. Dissipative corrections to the VA-predicted evolution are presented below, in a numerical form.

## 2. Generation of DMDS in random media

DMDS can be produced by a variety of DM maps, including ones with zero, normal, and anomalous average GVD [56], as shown in Figs. 1(a). The random component of the dispersion, which may be present in real fibers, is included in Fig. 1(b).

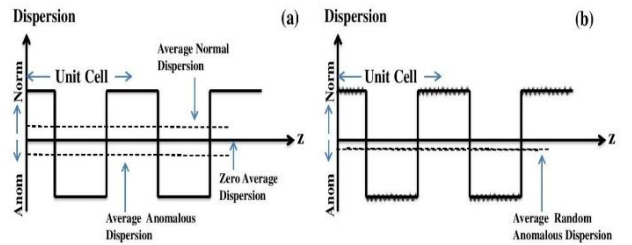

**Figure 1:** Different types of symmetric two-stage DM maps. Panels (a) represent the maps with the zero, normal, and anomalous average dispersion, while panel (b) depicts the anomalous dispersion with random dispersion added to each fiber segment.

A stable DMDS soliton can be obtained from VA-predicted evolution equations (i.e., Eq. 4 and 5) by the introduction of the gain with a proper strength. In particular, direct simulations initiated by the input predicted by the VA for the dispersion map from Fig. 1(b) demonstrate that the VA-predicted width and amplitude keep the initial values in the course of the simulations (not shown here in detail). However, this particular VA-based approach cannot capture the characteristic breathing behaviour of the DM soliton. To produce these dynamical features, one may follow the VA approach adopted in Ref. [51]. Alternatively, VA-predicted equations (4) and (5) can be solved, in congugation with a suitable numerical method, for the analysis of the formation of stable solitons. Here we analyze the generation and dynamics of DMDS through direct numerical solution of the governing Eq. (1).

In the numerical simulations presented below, each normal- and anomalous-GVD segment includes a 3% random-dispersion addition on top of the constant dispersion, the average GVD being anomalous. This version of the randomly affected DM map produces generic results, as shown by comparison with results of additional simulations.

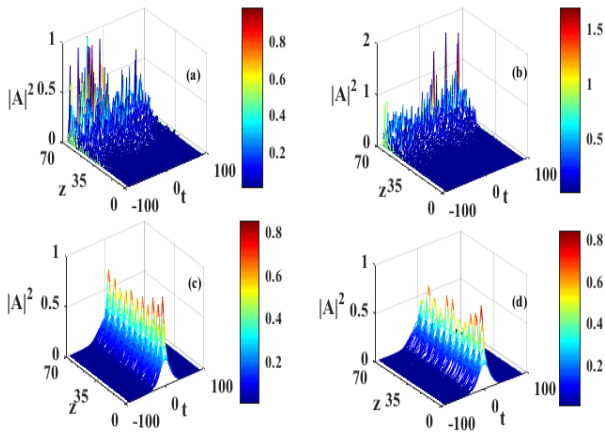

**Figure 2**: The evolution of the pulse under (a) zero average dispersion; (b) average normal dispersion; and (c, d) average anomalous dispersion but (d) includes randomness in each fiber segment. Here, parameters in Eq. (1) are $K= 0.01$, $\gamma= 0.01$, $\nu=0.01$ and $d=0.05$. The excess gain $\Delta g$ is $1.35 \times 10^{-6}$ and $1.47 \times 10^{-6}$ in (c) and (d), respectively. No stable solitons are produced in cases (a) and (b), for $\Delta g$ ranging from $-2.78 \times 10^{-6}$ to $1.50 \times 10^{-6}$.

Naturally, producing DMDSs is easier with the average anomalous GVD, as shown in Figs. 2(c, d), while the same system with zero and normal GVD leads to blow-up, see Figs. 2(a, b). The generation of the DMDSs under the anomalous average GVD may be controlled by tuning the gain, see Figs. 2(c, d).

The influence of randomness is noticeable too in Figs. 2(c, d), inducing randomness in the evolution of the peak power [which is a relatively strong feature in (d)], but it does not cause instability of the pulses. The randomness of the dispersion in the fiber segment may modelled differently, assuming, in particular, (i) uniform distribution of random values of dispersion (UDR), (ii) uniform distribution of integer random values of dispersion (UDRI), and (iii) random distribution of random values of dispersion (RDR). These three randomness, i.e., UDR, UDRI and RDR, have been generated by MATLAB tools, namely, 'rand', 'randi' and 'randn' respectively. Among these, 'rand' (UDR) and 'randi' (UDRI) generate random values following the statistical 'uniform distribution', while 'randn' (RDR) follows the statistical 'normal distribution'. Stable DMDS can be found for all these types of the randomness, as shown in Fig. 3(a, b, c).

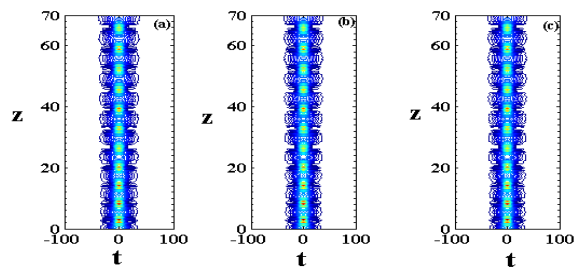

**Figure 3:** Contour plots display the robust evolution of DMDS for different types of the dispersion randomness in the system with the anomalous average GVD: (a) UDR; (b) UDRI; (c) RDR. Here, $K= 0.01$, $\gamma= 0.01$, $\nu=0.01$ and $d=0.05$. Here the net gain is $\Delta g = 1.35 \times 10^{-6}$ in (a), $1.59 \times 10^{-6}$ in (b), and $1.64 \times 10^{-6}$ in (c).

The typical breathing (expansion-contraction) rate of the DMDS is different for the different randomness patterns, being highest for RDR and lowest for UDR. The analysis following below is presented for the UDR pattern. In particular, the robustness if the breathing DMDS, i.e., persistence of their self-trapped shape, is shown by the amplitude-width plots in Fig. 4, both in the absence (a) and presence (b) of the random component in the DM map.

The nonlinear multiphoton absorption, gain dispersion and quintic nonlinearity all play a significant role in estimating the required gain for the creation of robust DMDS, as the results demonstrate for the system without and with random component in the DM map, see Figs. 5(a) and (b). Naturally, the increase of the gain dispersion, $d$, requires larger net gain to stabilize a DMDS. It is worthy to note that, as Fig. 5 demonstrates, that the interplay of the different effects in the model makes the necessary value of the gain slightly larger in

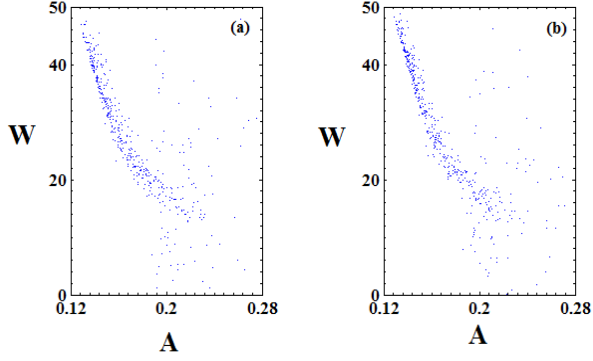

**Figure 4:** The amplitude-width plot for the DMDS under the anomalous average GVD, without (a) and with (b) random dispersion (of the UDR type), respectively. Fixed parameters are $K = 0.01$, $\nu = 0.01$, $d = 0.05$ and $\gamma = 0.01$. The excess linear gain is $\Delta g = 1.37 \times 10^{-6}$ in (a), and $1.45 \times 10^{-6}$ in (b).

the presence of both the TPA and 3PA than in the case when only the 3PA effect is present (the same feature is observed below in Fig. 6). In real fiber lasers, it may be possible that only one of these multiphoton-absorption effects is effectively present, or both are acting simultaneously [53, 54]. It is observed that for random dispersion map the net gain increases keeping the effect of multiphoton absorption unaltered (Figure 5b)

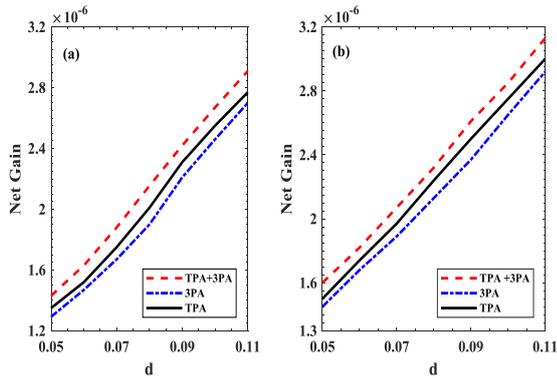

**Figure 5:** The variation of the net gain, $\Delta g$, required to generate the stable DMDS under the anomalous average dispersion, in the presence of gain dispersion $d$ and TPA or 3PA, or both these factors. Panels (a) and (b) display the necessary net gain in the absence and presence of the random-dispersion component. Other parameters are $K = 0.01$, $\gamma = 0.01$, $\nu = 0.01$.

Likewise, Fig. 6 demonstrates that the quintic nonlinearity controls the net gain required to stabilize the DMDS both for non-random and random DM maps, the increase of the quintic coefficient leading to conspicuous growth of the necessary net gain. The gains mentioned in the figure captions are the minimum gains required for the support and propagation of DMDS. An excessive gain leads to the blow up of the system.

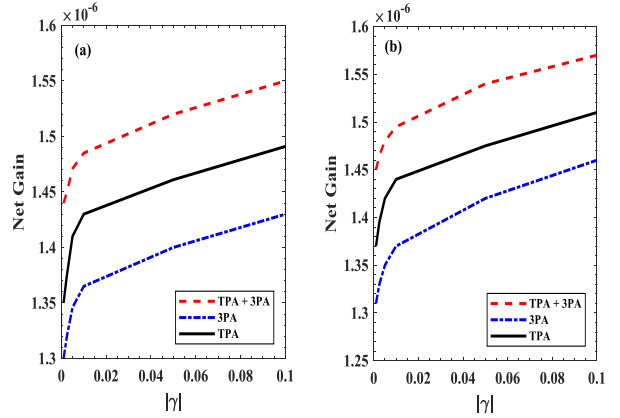

**Figure 6:** The variation of net gain $\Delta g$ required to generate the DMDS under the anomalous average dispersion, for different valued of the quintic-nonlinearity coefficient in Eq. (1), $|\gamma|$, in the presence of the TPA or 3PA, or both these factors. Panels (a) and (b) display the necessary net gain in the absence and presence of the random-dispersion component in the DM map. Other parameters are $K = 0.01$, $\nu = 0.01$, $d = 0.05$.

The effect of the random noise component on the formation of DMDSs is illustrated by Fig. 7. It is concluded that the soliton remains quite stable within the tolerance limit about 9%. Thus the considered mechanism is fairly robust for the implementation in fiber lasers.

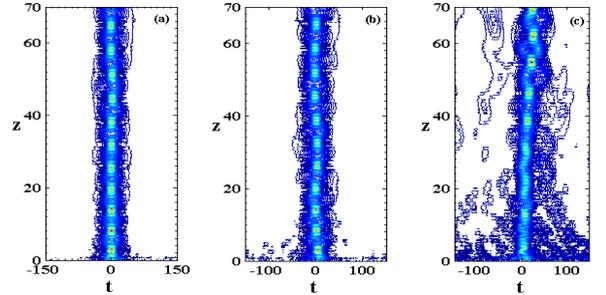

**Figure 7:** The evolution of the DMDSs under the anomalous average dispersion, in the presence of the random-dispersion component at levels 3% (a); 5% (b), and 9% (c). Other parameters are $K = 0.01$, $\nu = 0.01$, $d = 0.05$ and $\gamma = 0.01$.

The presence of the multiphoton absorption may induce the generation of multiple solitons, due to peak-clamping effect [58] or shaping of dispersive waves in the form of pulses. The increase of the gain by a large factor initiates creation of new pulses from the sidebands, which can eventually grow into solitonic pulses [Fig. 8]. The initial soliton may [Fig. 8(a)] or may not split [Fig. 8 (b), (c)], depending on the gain. Further increase of the gain may lead to blowup, as the pulse energy becomes too high for the system to sustain it.

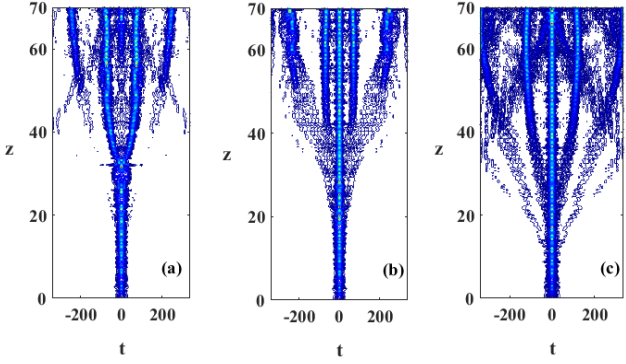

**Figure 8:** The generation of multiple dissipative solitons in the case of anomalous average GVD, in the presence of multiphoton absorption, for different values of the gain: (a) $3.45 \times 10^{-6}$, (b) $3.93 \times 10^{-6}$, (c) $4.47 \times 10^{-6}$. Other parameters are K= 0.01, ν=0.01, d= 0.05 and γ= 0.01.

The experimental verification of the present theoretical work can be performed using an experimental setup including a dispersion-managed fiber laser as the major component. An appropriate experimental model can be developed following the one used in Ref. [59], wherein the fiber laser ring cavity consists of an Erbium-doped fiber, a standard single-mode fiber, and a dispersion-compensation fiber, with positive, negative, and positive GVD, respectively. A fiber Raman laser source can be used as a pump in this setting. Additionally, the experimental model presented in Ref. [60] can be used too. Since our theoretical model considers randomness in the fiber segment, a lower-grade fiber segments may be incorporated.

Wavelength tuning of the DS fiber laser is an issue of fundamental interest due to various potential applications to spectroscopy and biomedical research. Saturable absorbers based on photonic topological insulators can passively Q-switch an erbium-doped fiber laser, producing high-energy ultra-broadband tunable pulses [61]. In the present model DMDS can be generated, in a given range of wavelengths, by adjusting the gain. A typical example of wavelength tuning for the DMDS shows (in Fig. 9) the corresponding variation in the required gain, which is necessary to stabilize the DMDS.

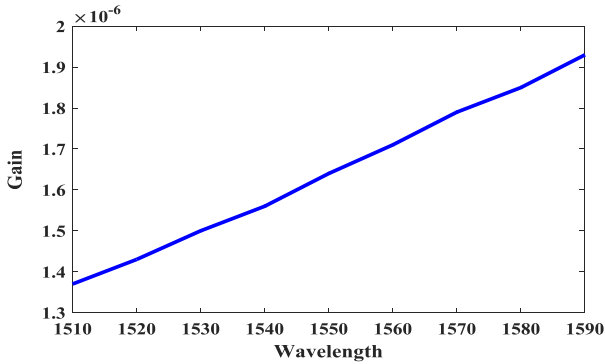

**Figure 9:** Gain required for the generation of tuneable DMDS in the fiber laser. Other parameters are K= 0.01, ν=0.01, d= 0.05 and γ= 0.01.

## 3. Interaction and switching of dissipative solitons

Whether it is a fiber-optic telecomm link (where soliton pulses are closely spaced) or a fiber laser, interactions between solitons are inevitable. While the self-phase modulation is important for the formation of solitons, the cross-phase modulation (XPM) strongly affects soliton interactions. In particular, XPM may lead to the creation of an induced soliton from two orthogonal polarization components, as first experimentally observed in a passively mode-locked fiber ring laser with a birefringence cavity [62]. The induced soliton generated by XPM of the two orthogonal polarization components in the birefringence laser may experience a frequency shift imposed by the primary (inducing) soliton.

The interaction between DMDSs is controlled by the separation and phase shift between them. In the current section, we study the interactions in the model with at least 3% dispersion randomness, unless it is stated otherwise. First, following the general principles of the soliton theory [63], DMDSs with zero phase shift, $\Delta\phi = 0$, attract each other. Accordingly, systematic simulations of the co-evolution of pairs of in-phase DMDSs reveal, with gradual increase of the initial separation between them $T_0$, merger into a single pulse [Fig. 10(a)], an effective rebound into a pair of far separated ones, due to the passage through each other of the original solitons accelerated by the mutual attraction [Fig. 10(b)], generation of two original pulses by two primary ones merging into one [Fig. 10(c)], which may be considered as incomplete mutual passage [cf. Fig. 10(b)], or no conspicuous interaction, if the initial separation exceeds $T_0 = 75$ [Fig. 10(d)], for parameter values fixed in the caption to Fig. 8.

Different scenarios of the interaction are observed for a non-zero initial phase shift $\Delta\phi$ between the two DMDSs, see Figs. 11 and 12. First, as explained in [64], $\Delta\phi \neq 0$ breaks the symmetry of the soliton-soliton collision. Indeed, the superposition of two phase-shifted solitons includes a wave-field factor, as a function of temporal distance $T$ between them:

$$F(T) = \cos(2\omega T + \Delta\phi), \qquad (6)$$

assuming that the mutual attraction accelerates the solitons, lending them inner frequency shifts $\pm\omega$. Thus, Eq. (6) gives rise to an offset of their *phase center*, $\Delta T = -\Delta\phi/\omega$, with respect to the *amplitude center* of the two-soliton superposition. The corresponding symmetry breaking is clearly observed in most cases displayed in Figs. 11(a, c) and 12(a, c). Nevertheless, it is worthy to note that the symmetry breaking is virtually absent in the case shown in Fig. 12(b), with $\Delta\phi = \pi/3$, which may be explained by proximity of this case to the orthogonality between the colliding solitons ($\Delta\phi = \pi/2$).

Further, it is also observed in Figs. 11 and 12 that the pairwise interaction between all DMDSs with $\Delta\phi \neq 0$ is always attractive, leading to their merger in a single pulse (unless the initial separation is very large), which is accompanied, in most cases, by ejection of an additional radiation jet. This difference from the outcomes of the collisions with $\Delta\phi = 0$ is explained by the fact that, after the initial fusion of the attracting solitons, the nonzero phase shift causes stirring of the wave field and prevents the re-splitting of the well-defined pulses, cf. Figs. 10(b, c).

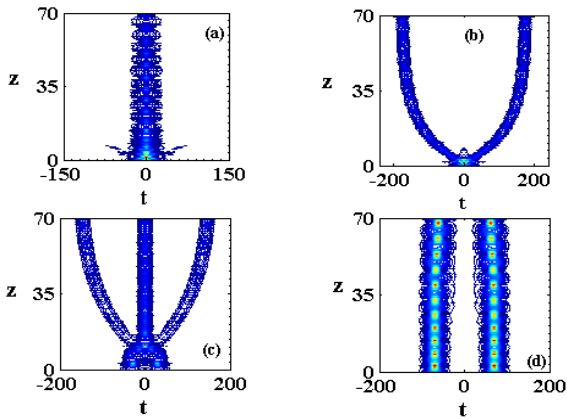

**Figure 10:** Interactions between two in-phase DMDSs for different initial separations ($T_0$) between them: (a) $T_0 = 5$, (b) $T_0 = 13$, (c) $T_0 = 32$, and (d) $T_0 = 75$. Other parameters are $K= 0.01$, $v=0.01$, $d= 0.05$ and $\gamma= 0.01$. The net gain is $\Delta g = 1.45 \times 10^{-6}$.

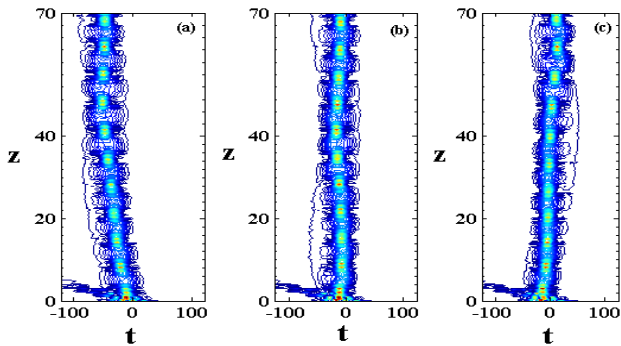

**Figure 11:** Outcomes of the interaction between two DMDSs (merger) for $T_0 = 10$ (a), $T_0 = 15$ (b), $T_0 = 20$ (c), with the initial phase difference $\Delta\phi = \pi/10$ between them. Other parameters are $K= 0.01$, $v=0.01$, $d= 0.05$ and $\gamma= 0.01$.

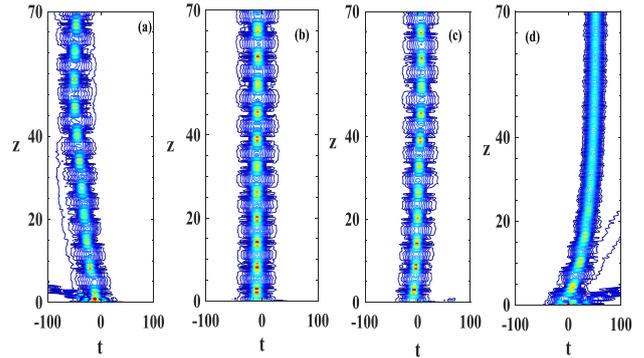

**Figure 12:** The same as in Fig. 9, but for the interaction of two DMDSs with fixed temporal separation, $T_0 = 10$, and different phase shifts: (a) $\Delta\phi = \pi/10$, (b) $\Delta\phi = \pi/3$, (c) $\Delta\phi = \pi/2$, (d) $\Delta\phi = 2\pi/3$.

Due to the action of the effective friction, induced by the spectral filtering [$\sim d$ in Eq. (1)], the originally moving pulses, produced by the collisions, come to the halt. Then, it is relevant to measure the total shift from the central position acquired by the merged soliton. The shift is plotted in Figs. 13 and 14, severally, as a function if the initial temporal separation, $T_0$, between the colliding DMDSs for a fixed phase difference between them, $\Delta\phi = \pi/10$, and as a function of $\Delta\phi$ for fixed $T_0 = 10$ [in the latter case, the colliding solitons quickly merge into the single pulse, see Fig. 11(a)]. The sign-changing form of the dependences observed in Figs. 13 and 14 is naturally explained by the cosinusoidal dependence in Eq. (6).

Lastly, the comparison of Figs. 13 and 14 demonstrates that the eventual shift of the merged pulse is more sensitive to the initial temporal separation between the colliding solitons, $T_0$, than to the phase difference between them, $\Delta\phi$. This conclusion may be explained by the fact that $T_0$ determines frequency shift $\omega$ in Eq. (6), through the mutual attraction between the initial solitons.

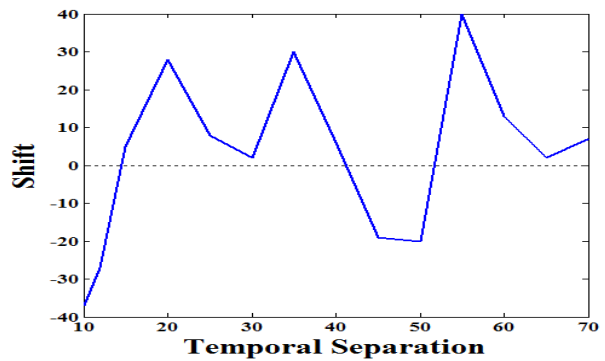

**Figure 13:** The final temporal shift of the pulse produced by the merger of two initial DMDSs with a fixed phase difference $\pi/10$, vs. the initial temporal distance $T_0$ between them. Other parameters are $K= 0.01$, $v=0.01$, $d= 0.05$ and $\gamma= 0.01$.

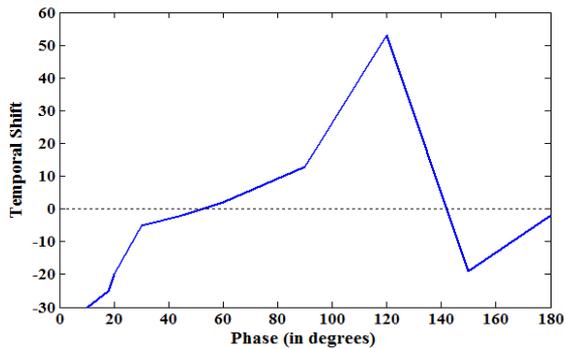

**Figure 14:** The same as in Fig. 11, but for the final shift vs. the phase shift between the colliding DMDSs, with fixed temporal distance between them, $T_0 = 10$.

At this point a comparison between Ref. [54] and the present study is in order. The first and foremost difference between the two systems lies in the nature of GVD. While Ref. [54] makes use of the conventional anomalous GVD, the current study uses the dispersion-management setting, which is based on the dispersion map with alternate anomalous and normal GVD. In comparison to usual solitons (in settings with anomalous GVD), DM solitons with zero or normal average GVD have a greater energy, better stability, larger signal-to-noise ratio, and weaker Gordon–Haus–Elgin timing jitter. All these feature fundamentally improve the system's performance. Although both Ref. [54] and the present work include random fluctuations in GVD (in the single fiber or fiber segments, respectively) and higher-order effects, values of parameters (the gain and others) required for the stabilization of the soliton are different. The interaction behaviour is also significantly different here and in Ref. [54].

## 4. Conclusion

We have developed systematic simulations for the generation of stable dispersion-managed dissipative solitons in the fiber laser with the cubic-quintic nonlinearity, gain dispersion (spectral filtering) and multiphoton absorption. The model's DM map includes a random dispersion component. The DMDSs (dispersion-managed dissipative solitons) exhibit the breather-like dynamics, remaining robust up to a certain (actually, quite high) level of noise. The interaction between in-phase DMDSs leads to their merger, effective rebound, or transformation of the three solitons into two. With non-zero phase difference between the solitons, the interaction leads to the merger, which is accompanied by the symmetry breaking, in the form of the shift of the merged pulse to the left or right. The latter effect is more sensitive to the initial temporal separation between the merging solitons than to their phase difference. These findings may be significant for the design of dispersion-managed fiber soliton lasers.

**Acknowledgements**

G. S. Parmar is grateful to Thapar Institute of Engineering and Technology, Patiala for providing financial support through Teaching Associateship. S. Jana would like to thank Ministry of Electronics & Information Technology, MeitY, Govt. of India for awarding "Young Faculty Research Fellowship under Visvesvaraya PhD Scheme" (Ref: MLA / MUM / GA / 10(37) B).